# COVID-19 societal response captured by seismic noise in China and Italy

Han Xiao, Zachary Cohen Eilon, Chen Ji and Toshiro Tanimoto

Department of Earth Science and Earth Research Institute, University of California, Santa Barbara, CA, USA

**Abstract**

Seismic noise with frequencies above 1 Hz is often called 'cultural noise' and is generally correlated quite well with human activities. Recently, cities in mainland China and Italy imposed lockdown restrictions in response to COVID-19, which gave us an unprecedented opportunity to study the relationship between seismic noise above 1 Hz and human activities. Using seismic records from stations in China and Italy, we show that seismic noise above 1 Hz was primarily generated by the local transportation systems. The lockdown of the cities and the imposition of travel restrictions led to a ~4-12 dB energy decrease in seismic noise in mainland China. Data also show that different Chinese cities experienced distinct periods of diminished cultural noise, related to differences in local response to the epidemic. In contrast, there was only ~1-6 dB energy decrease of seismic noise in Italy, after the country was put under a lockdown. The noise data indicate that traffic flow did not decrease as much in Italy, but show how different cities reacted distinctly to the lockdown conditions.

**Introduction**

Seismic background noise observed at frequencies above 1 Hz primarily consists of cultural noise, which is generated by human activities (Stutzmann *et al.*, 2000; McNamara and Buland, 2004; Groos and Ritter, 2009; Green *et al.*, 2017) such as trains (Chen *et al.*, 2004; Sheen *et al.*, 2009; Ribes-Llario *et al.*, 2017), road traffic (Hao and Ang, 1998; Coward *et al.*, 2003) and airports (Meng and Ben-Zion, 2018). These noise sources are now known to be useful for studying the subsurface structures (Nakata *et al.*, 2011; Riahi and Gerstoft, 2015; Quiros *et al.*, 2016; Ajo-Franklin *et al.*, 2019).

The outbreak of the novel coronavirus SARS-CoV-2 disease (hereafter: COVID-19) was first reported in Wuhan, Hubei, China, in December 2019 (Andersen *et al.*, 2020). In early to mid-January 2020, the virus spread to other Chinese provinces, facilitated by increased travel during the Chinese Lunar New Year. With Wuhan being a major rail transport hub in China, the virus quickly spread throughout the country. On 23 January 2020, Wuhan and other cities of Hubei province imposed a lockdown in an effort to quarantine the epicenter of the COVID-19 outbreak (Lu, 2020). All public transportation except for emergency and supply vehicles were suspended. A total of 12 other counties to prefecture-level cities in Hubei, including Enshi, the location of one of the seismic stations used in this study, were placed on travel restrictions by the end of 24 January 2020 (Table 1). The World Health Organization (WHO) declared the outbreak to be a Public Health Emergency of International Concern on 30 January 2020. By chance, this outbreak closely coincided with the Chinese Lunar New year of 2020, and the Chinese government utilized this coincidence to facilitate lockdown logistics, essentially extending the traditional week-long national holiday for several months. In Italy, following the rapid expansion of an outbreak of COVID-19 cases in the north of the country in late February 2020, the Italian government imposed a lockdown on many of its Northern provinces on 8 March 2020. The lockdown restricted all movement except for work, health circumstances and essential activities. On the evening of 9 March, quarantine measures were extended to the entire nation, becoming effective the next day. This quarantine included some important differences from the restrictions in China. For example, the lockdown did not apply to the public transportation system, including buses, railways, flights, and ferry services. People with self-declared travel exemptions were permitted to travel. On 11 March 2020, the WHO declared the outbreak a pandemic. This sudden declaration of a public health emergency provides an unusual dataset in which we can compare changes in seismic noise to the known timings of social orders.

In this study, we analyze continuous seismic time series from seismic stations in China and Italy (Fig. 1). We particularly focus on understanding the characteristics of cultural noise before and after the lockdowns. We show that the lockdown of cities in China led to a ~4-12 dB reduction in cultural noise. In contrast, there was only ~1-6 dB energy decrease of cultural noise in Italy, possibly due to ongoing traffic.

**Data and Methods**

We utilized data from the New China Digital Seismograph Network (NCDSN), in operation since 1992, with the network code IC. Data from IC were obtained from the Incorporated Research Institutions for Seismology (IRIS) Data Management Center (DMC;

www.iris.edu/dms) from 1 January 2000 to 15 April 2020. Those stations are: station IC.ENH, located in Enshi, Hubei province; station IC.MDJ, located in Mudanjiang, Heilongjiang province in the northeast region of China; station IC.BJT which is located in the northwest of Beijing, the capital of China; and station IC.QIZ which is located on Hainan island in Qiongzhou (Fig. 1a). We also analyzed data from the Italian National Seismic Network (INSN), with the network code IV. These data, including seismic stations IV.MILN in Milan, IV.MONC close to Torino, and IV.RMP ~20 km southeast of Rome, were obtained from the Italian National Institute of Geophysics and Volcanology (INGV; http://webservices.ingv.it) from 15 December 2019 to 15 April 2020 (Fig. 1b).

We analyzed broadband high-gain vertical seismograms (BHZ) and high-sample-rate high-gain broadband three-component seismograms (HHN, HHE, and HHZ), with sample rates of 20 Hz and 100 Hz, respectively. Ground acceleration records were retrieved by deconvolving the instrumental response from the original seismograms. All seismic data were divided into one-hour segments with overlapping time intervals of 50% (30 minutes). Each one-hour segment was detrended, tapered with a Hanning window, and the power spectral densities (PSD) were calculated. We did not remove earthquake signals from the time series, because the effects of earthquakes are limited to short time intervals and predominantly contain their most characteristic signals at lower frequencies. Thus they have extremely small overall effects on the estimation of cultural noise. We show noise power levels in units of decibels (dB) with respect to $10\log_{10}(\frac{\frac{m^2}{s^4}}{Hz})$ . We also performed a frequency-dependent polarization analysis for the high broadband three-component seismic data to determine the source directions of the cultural noise (Samson, 1983; Park *et al.*, 1987; Koper and Hawley, 2010).

**Results**

*Baseline seismic noise patterns at ENH*

We use a twenty-year-long seismic record from station IC.ENH, located in Enshi, Hubei province, China, to establish important baselines and patterns in the cultural noise, illustrating the detailed ways in which this signal is related to societal behavior. The vertical PSD for this station is shown in Figure 2. Enshi is a county-level city, about 460 km west of Wuhan, with a metro area population of 0.587 Million (https://www.macrotrends.net/cities/23612/enshi/population). Typical secondary microseism peaks are distinct in this figure with an approximate frequency band of 0.15-0.5 Hz. Another distinct peak can be identified at high frequencies, approximately in the 1.5-8 Hz frequency band. As we will show, this peak is caused by cultural noise.

There is a clear increasing trend in seismic noise in Enshi between 2000 and 2020 which shows a good correlation with the local economic growth and the number of civil motor vehicles (Figs 2b and 2c). Black arrows in Figure 2a show the timings of the Chinese Lunar New Year since 2000, correlating well with an annual lull in cultural noise due to decreased traffic flow and closure of factories. The red arrow in Figure 2a indicates the time that Enshi came under lockdown, resulting in a sudden decrease of cultural noise. We explore the effects of this lockdown by examining vertical component PSDs for the first three months of the year, comparing vertical PSDs between 2018 (Fig. 3a) and 2020 (Fig. 4a).

The 2018 data at this site provide information on various aspects of social life in a regular year, including diurnal variations, a holiday effect, and seasonal variations (Figs 3a, 5a and 5b). We find that two distinct cultural noise spectral peaks are evident, with approximate frequency bands of 1.5–8 Hz and 10-20 Hz. Diurnal variations are strong; there is higher seismic noise during the daytime than at night. The largest change in noise patterns is correlated well with the national holiday of the Chinese Lunar New Year, which typically starts one day before the New Year day and ends five days after. The noise lull is the result of cessation in typical travel and industrial activity. In 2018, this is evident from February 15 to February 21 (Julian days 46 to 52). During this period there is, on average, a 10 dB noise drop in the frequency band 1.5-8 Hz, compared to the background level. In detail, we show that the noise level started to drop 5-6 days before the holiday and did not fully recover until another two weeks after (Fig. 7). This frequency band also exhibits seasonal variations (Fig. 5a) which is longer strong noise signal in the summer than the winter corresponding to variation in daylight hours. We infer that 1.5-8 Hz noise is probably generated by the local factories, pedestrians (Alyamkin and Eremenko, 2011), and low-speed urban road traffic (Green *et al.*, 2017).

By contrast, the noise in the frequency band 10-20 Hz is relatively stable during the Chinese Lunar New Year. This persistence allows the noise source to be identified. Long (1971) found that moving vehicles on the freeway generate noise with peak frequency at about 10-20 Hz, and that this signal can be detected within 5-8 km. The station IC.ENH is located ~3 km from the freeway G50 to its southeast (Fig. 1a). The G50 is a 1900 km long and east-west bound expressway connecting Shanghai, China to the east and Chongqing, China to the west. While we are not yet able to retrieve the record of daily traffic-flow volume on the G50 near city Enshi, traffic data for a segment of this freeway named as Huanghuang east of Wuhan can be accessed (http://www.hhgs.org.cn/) and we use it as the representative of the general traffic flow on this road within Hubei province. The average daily traffic-flow volume on the freeway during the 2018 Chinese Lunar New Year was only 31% less than the daily volume on the regular days (Table 2). This suggests that the cultural noise peak in the frequency band 10-20 Hz can be explained by the high-speed cars on the freeway. This is further supported by the back azimuths of the polarization ellipsoids in Figure 3b. The source direction of cultural noise in the 10-20 Hz frequency band is from the southeast (back azimuths are ~ 120° – 150°) during the daytime, indicating the noise arrives from the direction of the freeway. The northwest noise source during the night (Fig. 3c) is probably from the downtown area, which lies in that direction. This dataset also shows a period of freeway traffic control between 13 September 2018 and 26 October 2018, when freeway access was limited or cut off due to road repairs. Station IC.ENH exhibits clear 10-20 Hz noise energy reduction during this period (Fig. 5b).

Our findings agree with previous analyses of frequency bands typical for road traffic (Groos and Ritter, 2009; Boese et al., 2015; Chang et al., 2016; Green et al., 2017) and consistent with a correlation between the cars' noise frequency content and their speed: the peak frequency band for the low-speed urban road traffic is 1.5-8 Hz, and for the high-speed cars traveling on the freeway it is 10-20 Hz (Long, 1971). Overground rail transportation could generate the seismic noise above 30 Hz in the seismograms (Chen *et al.*, 2004; Boese *et al.*, 2015; Riahi and Gerstoft, 2015; Green *et al.*, 2017), and the Chinese Lunar New Year is the peak travel season for the railway stations. However, according to our observations, there is a 5 dB reduction in the frequency band 30-40 Hz during this period, suggesting that the noise generated by the rail does not dominate our seismic record, perhaps due to the high attenuation and relatively far distance (8 km).

*Comparison of living habits between China and Italy.*

These data also reveal the various living habits of people in different cities (Fig. 5). For station IC.ENH located in Enshi (Fig. 5a and 5b), China, as we mentioned before, we can see the lull caused by the Chinese Lunar New Year in January or February (Fig. 2). In general, people in Enshi show most activity from 8 a.m. to 5 p.m. during the winter. In the summer, people tend to be active for longer, from 7:00 a.m. to 6:00 p.m. There is a lunch-time lull year-round at noon. However, for the frequency band 10-20 Hz generated by the traveling cars on the freeway (Fig. 5b), the lunch-time lull is longer and continues for about two hours during the summer. This may reflect hot noon-time weather in the summer discouraging people from going outside. Interestingly, the Chinese data do not show clear weekly cycles; work and activity continue week-round. In Milan, Italy (Fig. 5c and 5d), people work longer during weekdays (from 5 a.m. to 11 p.m.) and show clear differences in behavior on weekdays compared to the weekend. We can also see the cultural noise lull caused by local holidays, such as Easter Monday (April 22), Liberation Day (April 25),

Ferragosto (August 15) and Christmas (December 25). The cultural noise in the frequency band 10-40 Hz lull caused by the 2019 Christmas holiday for station IV.MILN is ~1 dB at the daytime and ~2 dB at nighttime (Fig. 8b). Interestingly, there are four distinct sources in frequency bands 1-2 Hz, 3-8 Hz, 10-20 Hz, and 20-40 Hz for station IV.RMP which located in the ~20 km southeast of Rome (Fig. 8e). We estimate the first peak in 1-2 Hz (Fig. 9a), with a ~5 dB reduction during the 2019 Christmas period, is generated by pedestrians (Alyamkin and Eremenko, 2011), since we expect fewer pedestrians on the street during the holiday of Christmas. By comparison to the Chinese data, we believe the second peak at 3-8 Hz (Fig. 9b) is generated by the local factories and low-speed cars on the street; this band a ~3dB deduction at Christmas. Finally, the high-speed cars on the freeway and the trains are responsible for the frequency bands at 10-20 Hz and 20-40 Hz respectively, which exhibit a ~2 dB reduction during Christmas (Fig. 8f). However, there is no obvious energy diminishing during the Christmas holiday for station IV.MONC (Fig. 8d), in which the noise is mainly from the freeway nearby (Fig. 1b). These observations are consistent with our observations from China that there are still lots of traveling cars on the freeway during the Chinese Lunar New Year holiday. The peak frequency of cultural noise for IV.MONC generated by the traveling cars on the freeway is ~20-30 Hz rather than 10-20 Hz. We estimate that this difference may relate to the different local geological conditions or simply the different regulations of highway speed limit (80 km/hr on G50 in Enshi, Hubei province in China vs. 130 km/hr in Italy). We note that there is a remarkable difference in the level of cultural noise (1-40 Hz) between IV.MILN, IV.RMP and IV.MONC, as stations IV.MILN and IV.RMP exhibit much higher noise. We believe the reason is that stations IV.MILN and IV.RMP is in an urban area, whereas IV.MONC is in a mountainous area (Fig. 1b). Italy follows the European Summer Time annual Daylight Saving Time procedure setting the clocks forward one hour from standard time during the summer months. In 2019, summer time was from March 31 to October 27. Figures 5c and 5d show a clear time shift of cultural noise energy due to these clock changes. The abrupt time shift in the frequency band 10-40 Hz (Fig. 5d) at the time of clock changes is reflective of the fact that this energy is generated by public transportation, such as trains, and buses, which have a fixed schedule. We highlight how cultural noise data reflects nuances in societal behavior in order to illustrate how these data can provide a detailed proxy account of the societal COVID-19 response.

*Cultural noise changes in China*

In 2020 we observe a sharp decrease in cultural noise in (Figs 4a and 4d) which coincides with the time when the city of Enshi went under lockdown due to COVID-19 (Table 1), and the extended new year holiday. By comparison to historical data (Fig. 2), in the 1-8 Hz band the ~12 dB decrease was equivalent to the abrupt cessation of roughly 20 years' worth of urbanization and development activity. In Figure 4a, a weak peak still appears during the lockdown time in the frequency band around 10-20 Hz. This could be caused by the official vehicles and the supply vehicles on the road (Groos and Ritter, 2009; Boese *et al.*, 2015; Chang *et al.*, 2016; Green *et al.*, 2017). The energy of this peak increases steadily as more vehicles appear on the road from approximately Julian day 70 onwards. Directional analysis of this noise reveals that it mainly comes from the southeast even at night (Fig. 4b and 4c) where the national freeway is located (Fig. 1a). The systematic increase in traffic as the lockdown eased serves as a natural experiment that we can leverage to better understand the relationship between traffic and seismic noise. Seismic noise generated by pedestrians and local industry is usually in the frequency band 1-5 Hz (Alyamkin and Eremenko, 2011). On the one hand, the marked decrease of seismic energy in this frequency band during the lockdown period (Fig. 4a) reflects the many fewer pedestrians and cars on the street. On the other hand, the observation that 1-5 Hz social noise increased ~5 dB in a two week period from Julian days 71 to 85 suggests that Enshi started to gradually reopen much earlier than the official lift of the lockdown. Note that the increase in social noise at Enshi is correlated well with the traffic flow volume at the highway segment 600 km away (Fig. 4d). It suggests that this gradual reopening is a province-wide activity.

The lockdown of Hubei province came one day before the 2020 Chinese Lunar New Year (January 25), the most important festival in the country. To quantify the energy reduction in cultural noise caused by the coronavirus alone, we compared the daily cultural noise energy variation between 2018 and 2020 using the day of Chinese Lunar New Year as the reference time in Figure 7. It is worth noting that they show a similar pattern before the lockdown of the cities in the Hubei province. However, after the lockdown, the cultural noise energy in 2020 is much lower than in 2018. The average reduction was ~10 dB in the frequency band 1.5-8 Hz and ~12 dB in 10-20 Hz for the station IC.ENH (Fig. 7c).

We conducted a similar analysis at several other stations located within urban centers in China (Fig. 6 and Fig. 7). None of these regions came under direct lockdown. We again compared the vertical component power spectral densities between the years 2018 (Fig. 6, left) and 2020 (Fig. 6, right). We found that the peak frequencies of cultural noise appear to be different for different cities. This is probably due to the relative distances to the noise sources and the installation (environmental) conditions at different station sites (Trnkoczy *et al.*, 2012).

For station IC.MDJ, located in the northeast region of China, the cultural noise is seen in mainly two peaks, similarly to at IC.ENH. One spans the frequency band 5-10 Hz and the other the frequency band 10-30 Hz. The first peak seems to be consistent with the local road traffic, and the second peak is consistent with the noise by the nearby railway that is at a distance of about 3 km. A substantial change in noise is observed coincident with the lockdown of Hubei and Lunar New Year. The duration and magnitude of this noise change, when compared to the 2018 record, demonstrates that this change substantially exceeded the 'normal' variation due to the new year holiday, indicating that industry and civilians altered behavior in Mudanjiang in response to COVID-19 despite the lack of formal local lockdown. There was a ~3 dB reduction in the frequency band 5-10 Hz and ~4 dB in the frequency band 10-30 Hz (Fig. 7a). The lowest noise conditions persisted for ~20 days, followed by a slow return to normal noise levels over a further ~60 day period.

Station IC.BJT, located in Beijing, shows only the lower frequency cultural noise, in this station mostly at 2-5 Hz (Figs 6c and 6d). This may be related to site installation: this seismograph was installed in a deep tunnel, which might suppress high-frequency noise (McNamara and Buland, 2004). This 2-5 Hz cultural noise is likely to be generated by both the road traffic and by pedestrians (Alyamkin and Eremenko, 2011; Boese *et al.*, 2015; Green *et al.*, 2017). As with other stations, the COVID-19 effects produced a protracted lull in the cultural noise, with a reduction ~4 dB in 2-5 Hz (Fig. 7b). At this station the duration of the noise reduction was longer, more than 81 days. The return to 'normal' cultural noise levels was much substantially more gradual than at other stations, with a slow increase in amplitudes from Julian day 30 and recovery to early-January noise levels at approximately Julian day 100 (76 days after the Hubei lockdown started). At time of writing (Julian day 102) noise levels at IC.BJT are still lower than the 'normal' background noise levels of 2018, implying a persistent alteration in traffic and social patterns from 'business as usual'.

Station IC.QIZ, located on Hainan island, which is famous for its tourism industry during the winter, shows a similar pattern to IC.ENH (Figs 6e and 6f), with a dominant cultural noise peak in the range 2-20 Hz that seems to include distinct sources in bands 2-8 Hz and 8-20 Hz. This is probably because both stations are closer to the freeways (unlike station IC.MDJ) (Fig. 1a). IC.QIZ manifests a similar noise variation to IC.MDJ, with a ~30 day lull, followed by a gradual return to 'normal' but still less than the background levels in 2018 over a further ~51 day period (Fig. 7d). The average reduction is ~10 dB in the frequency band 2-8 Hz and ~8 dB in the frequency band 10-20 Hz. Interestingly, the higher frequency (10-20 Hz) noise at this station seems to recover faster than the lower frequency (2-8 Hz) noise. If the former reflects high-speed vehicular traffic and the latter reflects pedestrian traffic, this staggered recovery may result from civilians feeling safe travelling in their own cars earlier than they feel comfortable walking around.

*Cultural noise changes in Italy during lockdown*

Italy was put under a dramatic lockdown (Table 1) as the coronavirus continued to spread in the country. Although it was one of the toughest responses implemented outside of China, their lockdown policy was less strict than China. As a result, we might expect traffic noise not to have decreased as sharply as we found in mainland China. We find only ~1 dB of energy decrease of cultural noise in the frequency band 10-40 Hz in IV.MILN and ~5 dB of decrease in IV.MONC after Italy declared its lockdown (Fig. 8). For station IV.RMP, the energy reduction was ~6 dB in 1-2 Hz (Fig. 9a), indicating many fewer pedestrians on the street. There is a ~4 dB reduction in the frequency band 3-8 Hz (Fig. 9b) and ~5 dB reduction in the frequency band 10-40 Hz (Fig. 8f), which implies the decrease in traffic-volume was less than the reduction in foot-traffic. Our observations are consistent with the local lockdown policies. The Italian authorities required that their schools, universities, theaters, cinemas, bars, and nightclubs must be closed. Religious gatherings, including funerals and weddings, and sporting events were suspended or postponed. Restaurants and bars were allowed to be open from 6 a.m. to 6 p.m., and shopping malls and markets could open on weekdays with a decreased density of patrons. Under such conditions, the cultural noise should be primarily generated by transportation systems. Lack of any decreasing seismic noise across the lockdown timing seems to corroborate the inference that the primary noise source was public transportation, which was not impacted by the lockdown (Pepe *et al.*, 2020). It appears that the continuous operation of the public transportation system maintained the persistently high level of cultural noise.

Despite this, a modest but significant decrease in noise level is observed at all Italian stations, from a period beginning at the official lockdown until at least Julian day 106 (time of writing). At stations with higher overall cultural noise (IV. MILN and IV.RMP), the pattern of noisy weekdays and less noisy weekends continues after the lockdown, although both shift to lower-noise than their pre-lockdown counterparts. In fact, for IV.RMP, near Rome, the post-lockdown week days are less noisy than even the quiet pre-lockdown weekends.

For station IV.MILN, the lowest noise energy appears in the first weekend after lockdown of the country, with the lowest noise conditions persisting just for one day. We also note that even the quietest post-lockdown day is not as quiet as the 2019 Christmas day. Since Julian day 74 we observe a slow increase in noise over a further 40 day period, perhaps indicating that civilians are increasingly willing to go outside in Milan. However, the noise levels have not yet reached pre-lockdown levels. For station IV. MONC, seismic noise reduced ~5dB following the lockdown over a period of 5 days. There is no clear trend of noise increase at this station, perhaps indicating a more strict maintenance of social distancing and stay-at-home behavior. Station IV.RMP, by contrast, recorded a near-immediate reduction in seismic noise over the few days following the lockdown, and actually manifests a gradually decreasing trend for the entire cultural noise frequency band 1-40 Hz. The decrease is particularly evident in the 1-2 Hz and 3-8 Hz period bands associated with pedestrians and local urban traffic (Fig. 9). This trend may imply that people in Rome are increasingly concerned by the COVID-19 pandemic and are adjusting their behavior to be more conservative.

**Conclusions**

Seismic records provide unique signals that can elucidate human activities on a large scale. In this paper, we examined variations in seismic noise between 1 Hz and 40 Hz, which provide proxy information on cultural behavior. In particular, we focused on the effects of governmental lockdowns and self-imposed behavioral alterations due to the outbreak of COVID-19 in mainland China and Italy. Using seismic records from stations in China and Italy, we show that the cultural noise in the range of about 2-40 Hz was primarily generated by the local transportation and population sources and study the living habits of local people by using seismic data. The lockdown of the cities and imposition of travel restrictions led to a ~4-12 dB energy decrease in cultural noise on the background of the noise energy in mainland China. According to our observations, different Chinese cities experienced distinct periods of diminished cultural noise, related to the differing local responses to the epidemic. A marked noise change was found even in cities that did not come under government-mandated quarantine. In contrast, there was only ~1-6 dB energy decrease of cultural noise after Italy was put under a total lockdown, due to continuous public transport. Italian cities seem to be responding differently in terms of social behavior as the lockdown continues.

**Data and Resources**

The data used in this study were collected from the Incorporated Research Institutions for Seismology (IRIS) Data Management Center (DMC; www.iris.edu/dms) and the Italian National Institute of Geophysics and Volcanology (INGV; http://webservices.ingv.it) in Italy using ObsPy Python package (Beyreuther et al., 2010). We used GMT (Wessel and Smith, 1991) to make many of the figures in this paper. Our data for seismic PSD, polarization results and traffic-flow volume on the freeway can be obtained from https://zenodo.org/record/3740214#.XojbDC2ZNE6.

**Acknowledgments**

The authors thank Mohan Pan, Kaelynn Rose, Scott Condon, and Brennan Brunsvik for comments which improved the manuscript substantially. This work was supported by grants from the Southern California Earthquake Center (SCEC #19037, #20072). SCEC is funded by NSF Cooperative Agreement EAR-1600087 & USGS Cooperative Agreement G17AC00047.

Han Xiao

Email: hanxiao@ucsb.edu

Address: Department of Earth Science and Earth Research Institute University of California, Santa Barbara 1006 Webb Hall Santa Barbara, California 93106

Zachary Cohen Eilon

Email: eilon@ucsb.edu

Address: Department of Earth Science and Earth Research Institute University of California, Santa Barbara 1006 Webb Hall Santa Barbara, California 93106

Chen Ji

Email: ji@geol.ucsb.edu

Address: Department of Earth Science and Earth Research Institute University of California, Santa Barbara 1006 Webb Hall Santa Barbara, California 93106

Toshiro Tanimoto

Email: toshirotanimoto@ucsb.edu

Address: Department of Earth Science and Earth Research Institute University of California, Santa Barbara 1006 Webb Hall Santa Barbara, California 93106

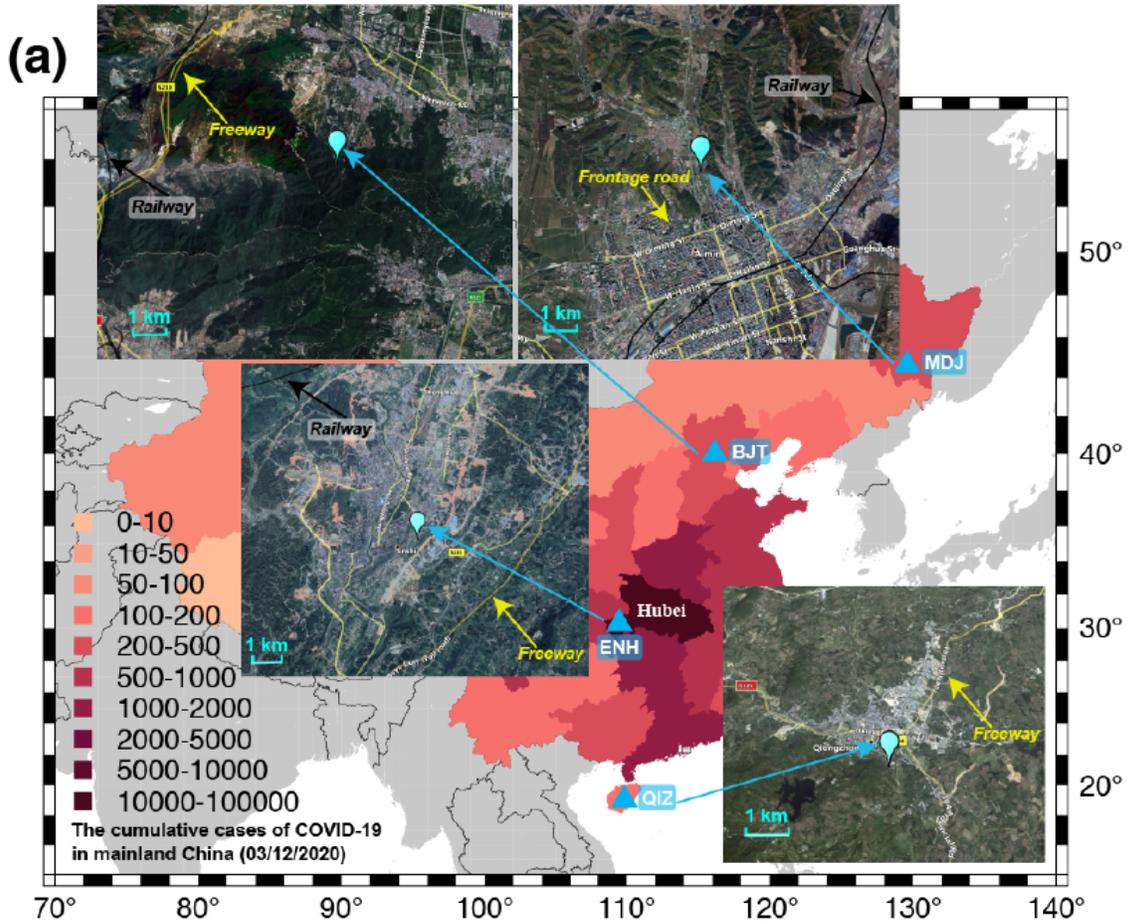
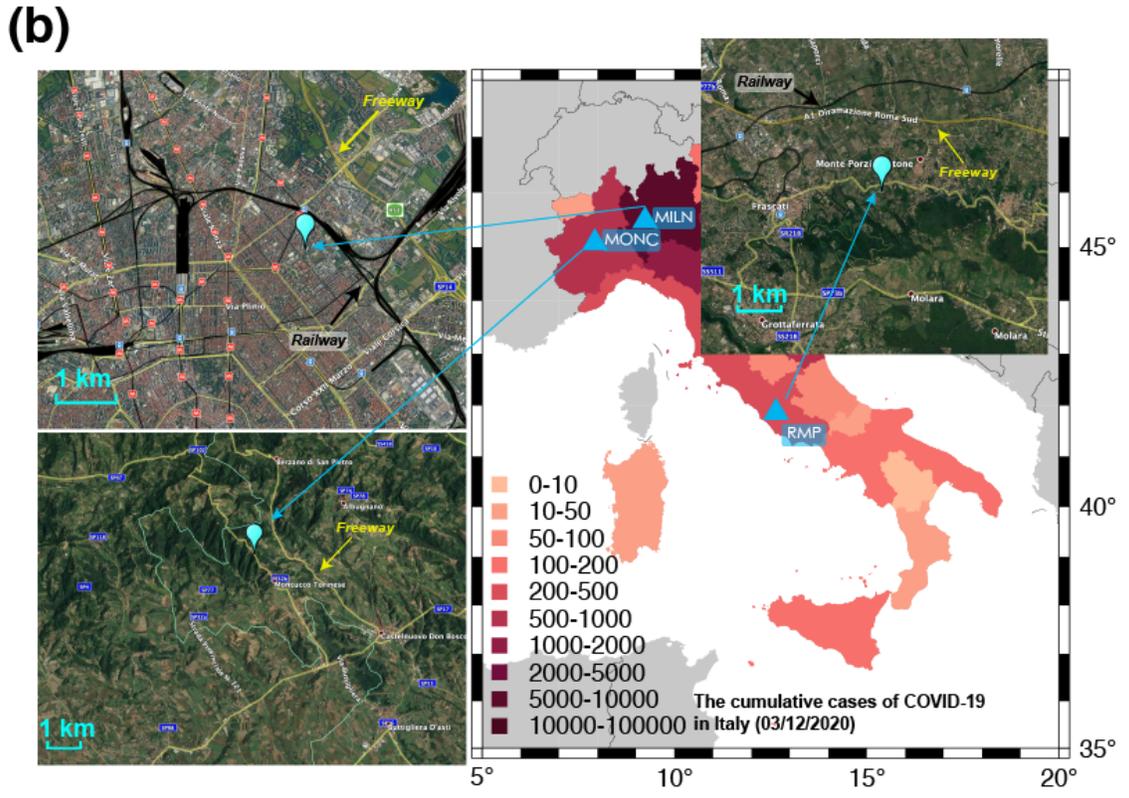

Figure 1. (a) Map of stations in China. The blue triangles indicate seismometer locations. The base image shows the cumulative cases of COVID-19 for different provinces in mainland China as of 12 March 2020, based on data from the World Health Organization. The insets show the seismometer locations in each urban environment. IC.MDJ is located close to a railway line and a major road. IC.BJT is also close to the freeway and railway line but the seismometer is deployed in the deep tunnel. IC.ENH is located north west of a major freeway and south east of the local urban center. IC.QIZ is on the Hainan island and is also close to the freeway to its north. (b) Same as (a) but for Italy, highlighting seismic stations IV.MILN in Milan, IV.MONC in the area of Torino, and IV.RMC near Rome. Inset maps show local maps: station IV.MILN was deployed in the urban area along the railway line and freeway. Station IV.MONC is in the mountainous area but close to the freeway. Station IV.RMP is in the suburbs 20 km southeast of Rome, Italy's capital city.

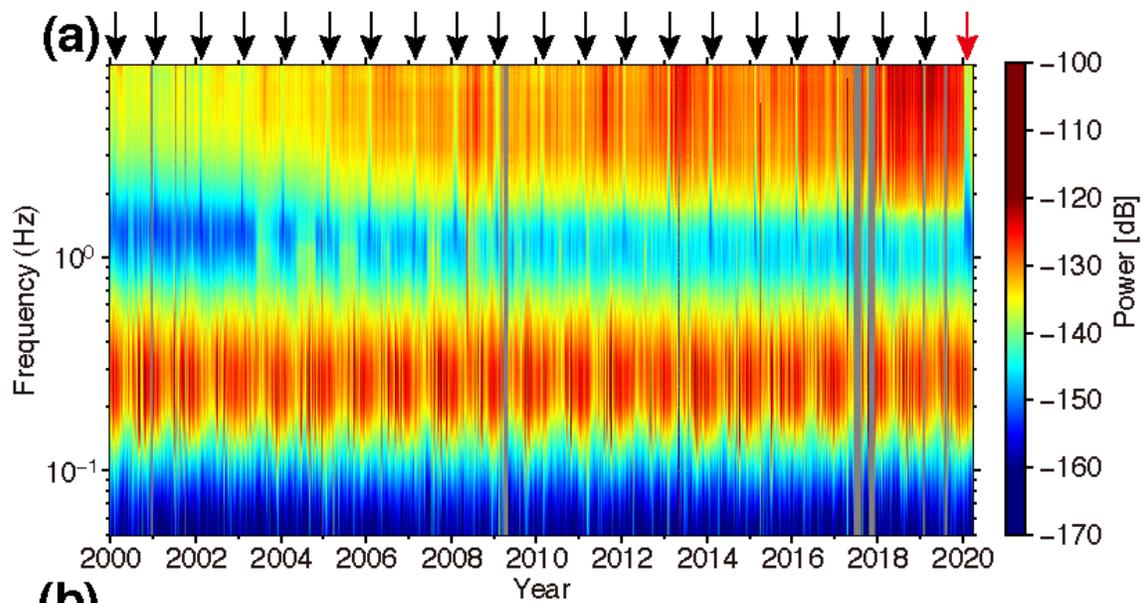
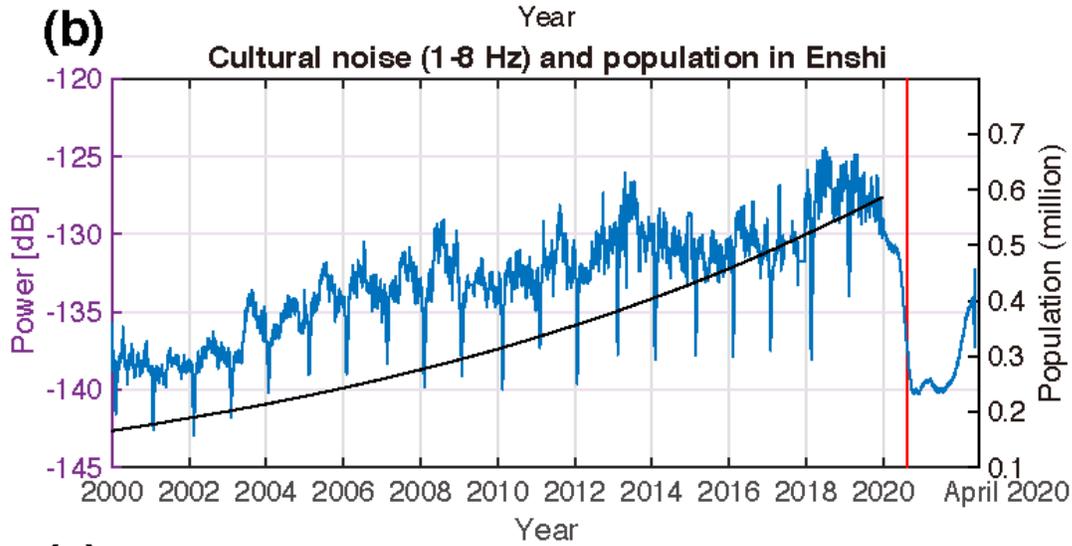
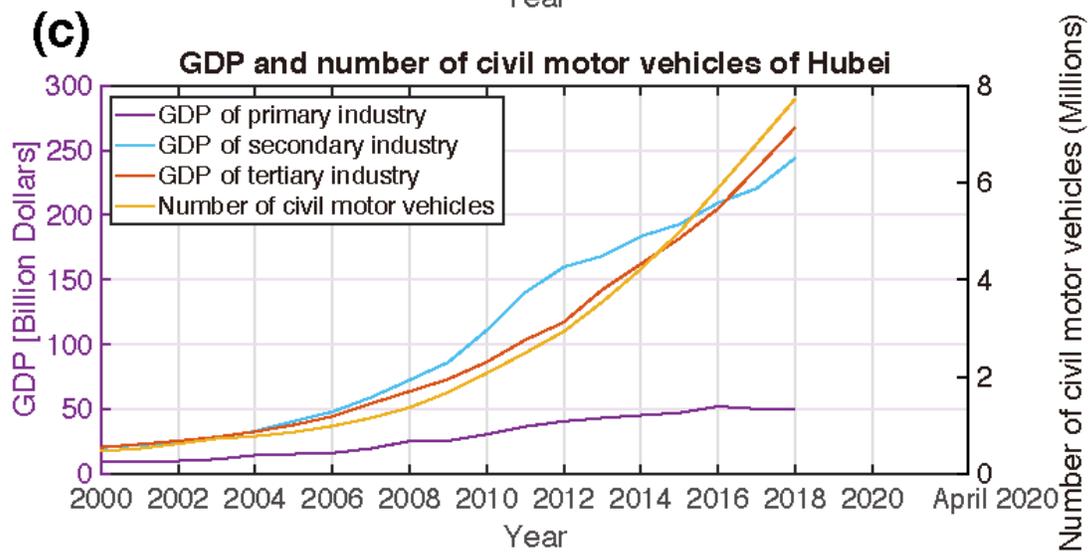

Figure 2. (a) Twenty years of power spectral density (PSDs) analysis for IC.ENH (Enshi, Hubei province) which has been operational since 20 September 1997. PSDs from the vertical component are shown in decibels relative to the ground acceleration with units of $10\log_{10}(\frac{\frac{m^2}{s^4}}{Hz})$. Black arrows indicate the timing of the Chinese Lunar New Year; the red arrow (the top-right location) indicates the time when the city went under lockdown due to COVID-19. (b) Twenty-year variation of cultural noise in the frequency band 1-8 Hz. The red line indicates the timing that Enshi went under lockdown due to COVID-19. The black line denotes the metro area population of Enshi during this period (c) GDP and the number of civil motor vehicles for the period 2000-2018 in Hubei province. Data are from the China National Bureau of Statistics (http://www.stats.gov.cn/).

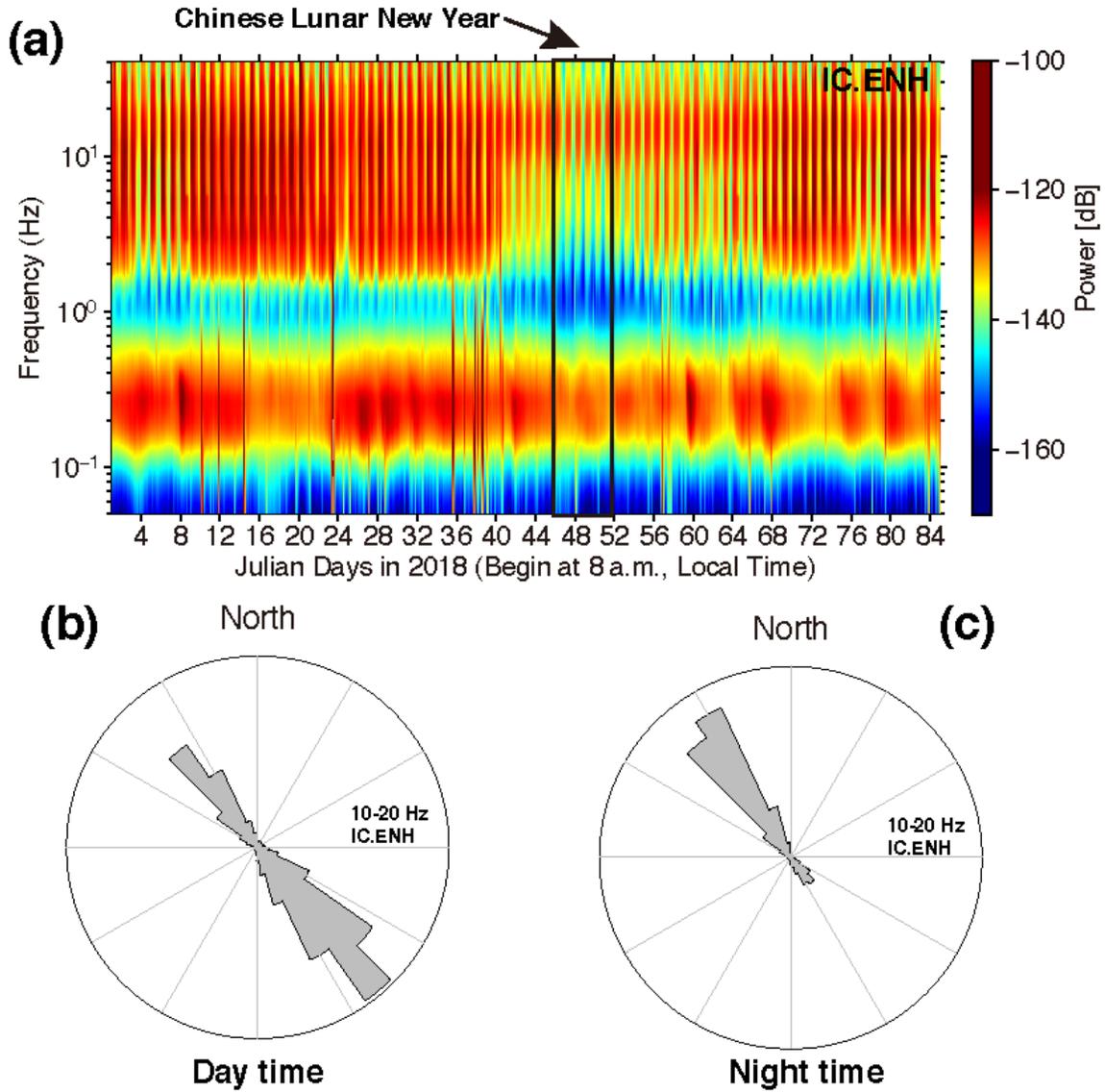

Figure 3. (a) Power spectral density of vertical component (HHZ) for the station IC.ENH in Enshi, Hubei province from Julian days 1 to 86 in 2018. PSDs from the vertical component are shown in decibels relative to the ground acceleration with units of $10\log_{10}(\frac{\frac{m^2}{s^4}}{Hz})$. The diurnal variations of cultural noise are obvious in the PSDs; seismic noise is higher during the day than the night. The black box indicates the period of the Chinese Lunar New Year (Julian days 46 to 52). Typically, the low cultural noise period is found during the Chinese Lunar New Year for a duration of about one week. (b)

Distribution of back azimuths for the frequency band 10-20 Hz at daytime (7:00 a.m. to 7:00 p.m., local time) estimated from one-year data in 2018 for IC.ENH. (c) Same with (b) but at nighttime (7:00 p.m. to 7:00 a.m., local time). The source direction of cultural noise in the 10-20 Hz frequency band is from the southeast at the daytime, and from the northwest at the nighttime.

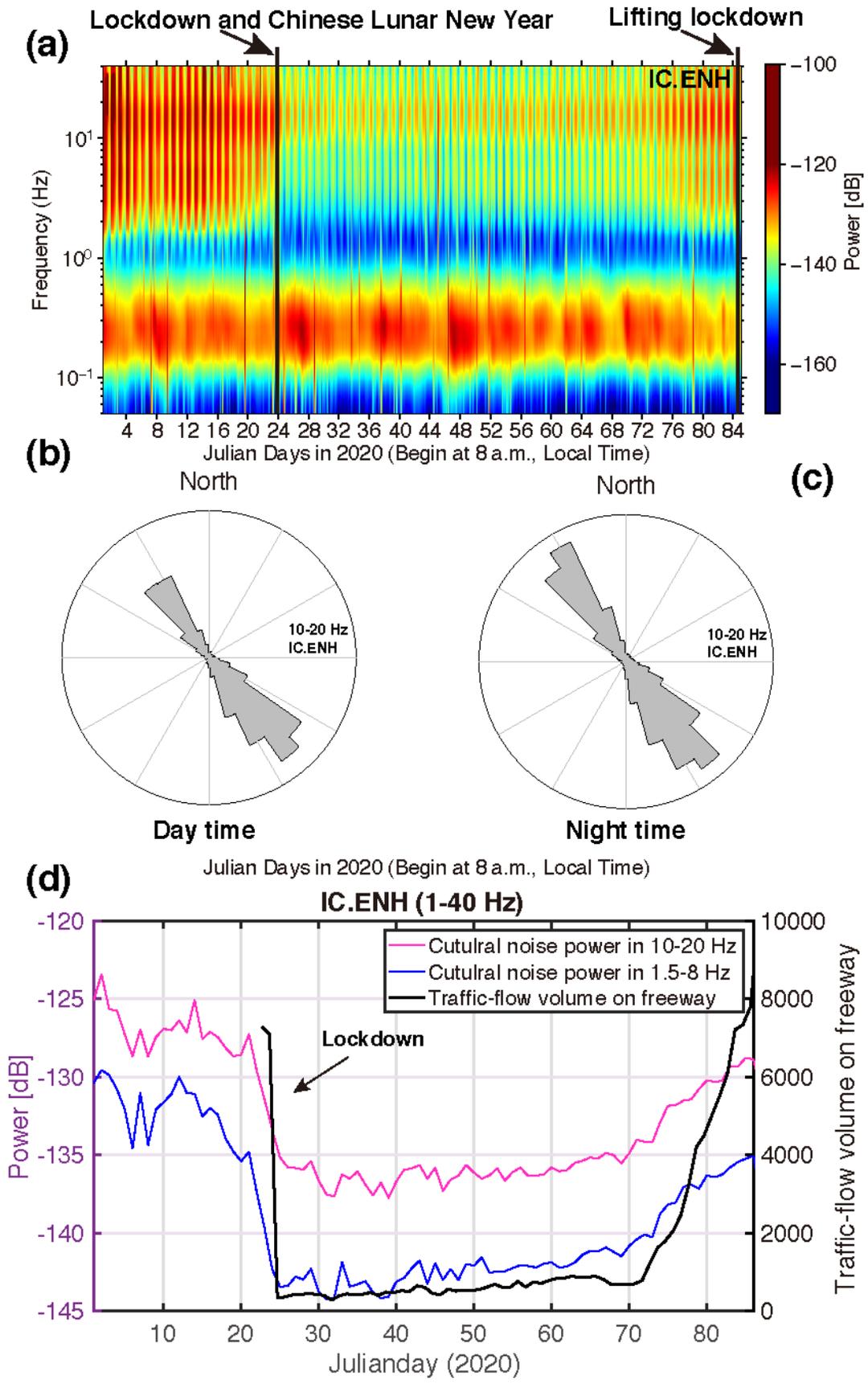

Figure 4. (a) The vertical component noise PSDs for the station IC.ENH in Enshi, Hubei province from Julian days 1 to 86 in 2020. The black line indicates the timing the city went under lockdown due to COVID-19. (b) The back azimuths of the polarization ellipsoids for the different frequencies as a function of time from Julian days 1 to 86 in 2020. (b) Distribution of back azimuths for the frequency band 10-20 Hz at daytime (7:00 a.m. to 7:00 p.m., local time) estimated from Julian days 24 to 86 when Enshi was under lockdown. (c) Same with (b) but at nighttime daytime (7:00 p.m. to 7:00 a.m., local time). (d) Comparison of cultural noise in the frequency band 10-20 Hz (red line) and 1.5-8 Hz (blue line) with the daily traffic-flow volume on the freeway, the traffic data is from the website http://www.hhgs.org.cn/.

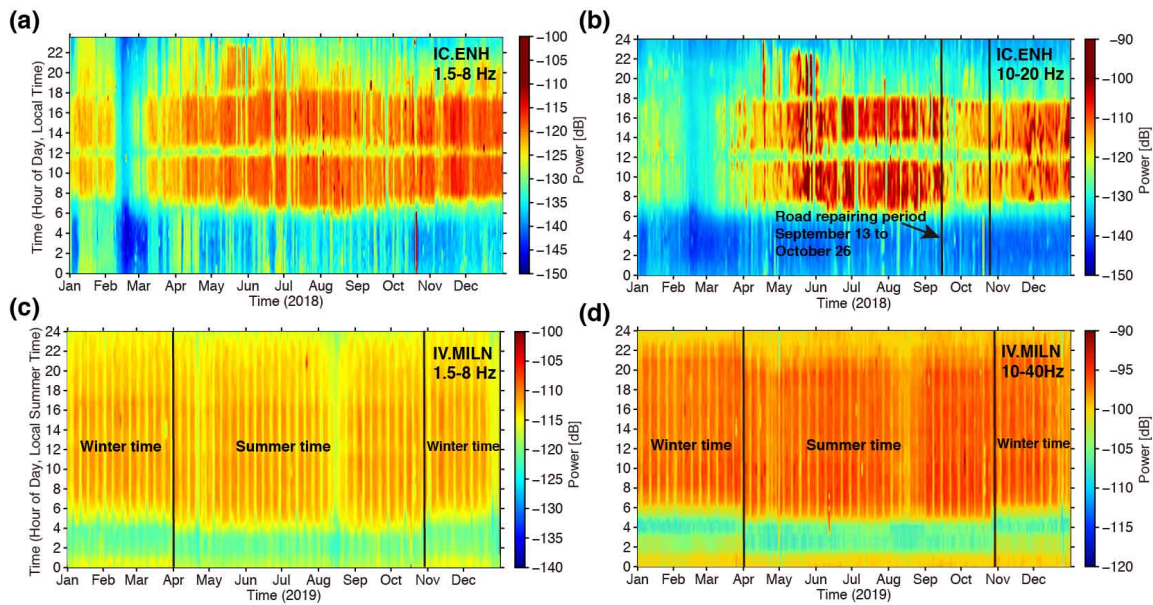

Figure 5. (a, b) The noise variations for the frequency band 1.5-8 Hz and 10-40 Hz in half-hour bins across the year of 2018 for the station IC.ENH in Enshi, Hubei province, China. The power is measured from the vertical component in decibels relative to the ground

acceleration with units of $10\log_{10}(\frac{\frac{m^2}{s^4}}{Hz})$. Black lines in (b) show the repairing period of the local freeway which is located in the southeast of seismic station IC.ENH. Note there is a constant lull at 12:00 local time. (c, d) Same with (a, b) but for the station IV.MILN in Milan, Lombardy province, Italy, in 2019. The black lines show the summer time in Italy, which is from March 31 to October 27 in 2019.

| Table 1 | | | |
|---|---|---|---|
| **City** | **Seismic station** | **The timing of the lockdown** | **The timing of the lifting lockdown** |
| Enshi (Hubei province, China) | IC.ENH | January 24, 2020 (Julian days 24 in 2020) | March 25, 2020 (Julian days 85 in 2020) |
| Beijing (China) | IC.BJT | No official declared lockdown | |
| Mudanjiang (Heilongjiang province, China) | IC.MDJ | No official declared lockdown | |
| Qiongzhou (Hainan province, China) | IC.QIZ | No official declared lockdown | |
| Milan (Lombardy province, Italy) | IV.MILN | March 8, 2020 (Julian days 68 in 2020) | Still in place at time of writing (April 19th) |
| Torino (Piedmont province, Italy) | IV.MONC | March 10, 2020 (Julian days 70 in 2020) | Still in place at time of writing (April 19th) |
| Rome (Lazio province, Italy) | IV.RMP | March 10, 2020 (Julian days 70 in 2020) | Still in place at time of writing (April 19th) |

| Table 2 | | |
|---|---|---|
| | **Average in 2018** | **During 2018 Chinese Lunar New Year (Julian days 46-52)** |
| Daily traffic-flow volume on freeway | 18950 vehicles | 13139 vehicles |

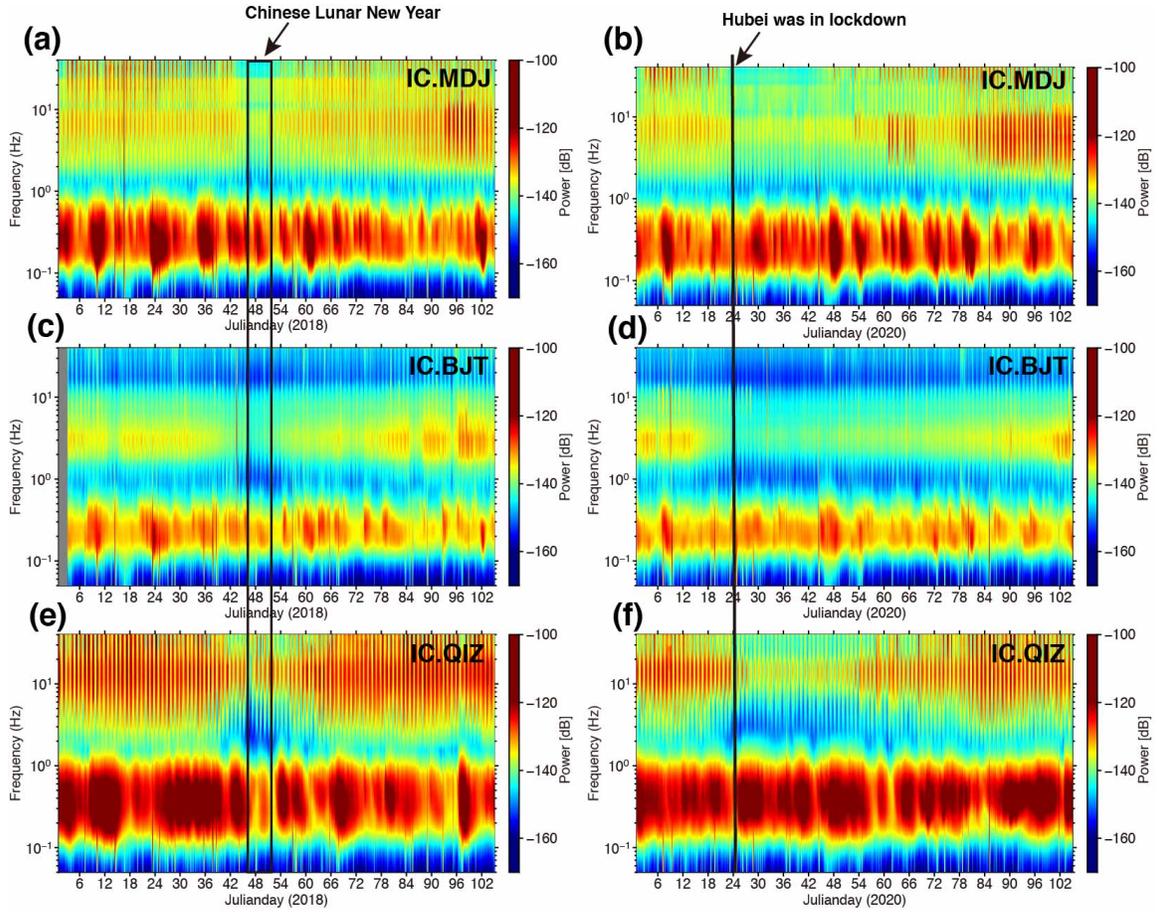

Figure 6. Comparison of the noise PSDs between 2018 and 2020 for the station (a, b) IC.MDJ (Mudanjiang), (c, d) IC.BJT (Beijing) and (e, f) IC.QIZ (Qiongzhou). The power is measured from the vertical component in decibels relative to the ground acceleration with units of $10\log_{10}(\frac{\frac{m^2}{s^4}}{Hz})$.

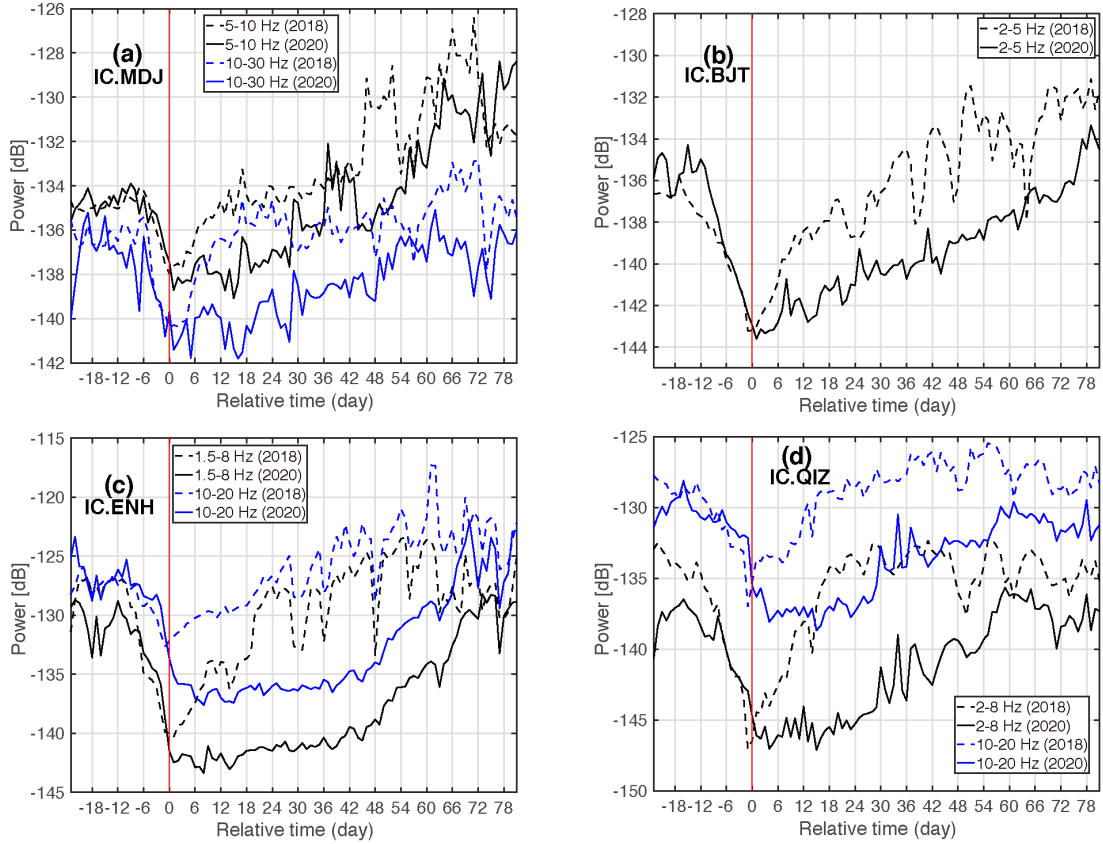

Figure 7. Comparison of the cultural noise energy daily variation between 2018 and 2020 for the station (a) IC.MDJ (Mudanjiang), (b) IC.BJT (Beijing), (c) IC.ENH (Enshi) and (d) IC.QIZ (Qiongzhou). The power is measured from the vertical component in decibels relative to the ground acceleration with units of $10\log_{10}(\frac{\frac{m^2}{s^4}}{Hz})$. The time is aligned with the day of the Chinese Lunar New Year which is indicated by the red line.

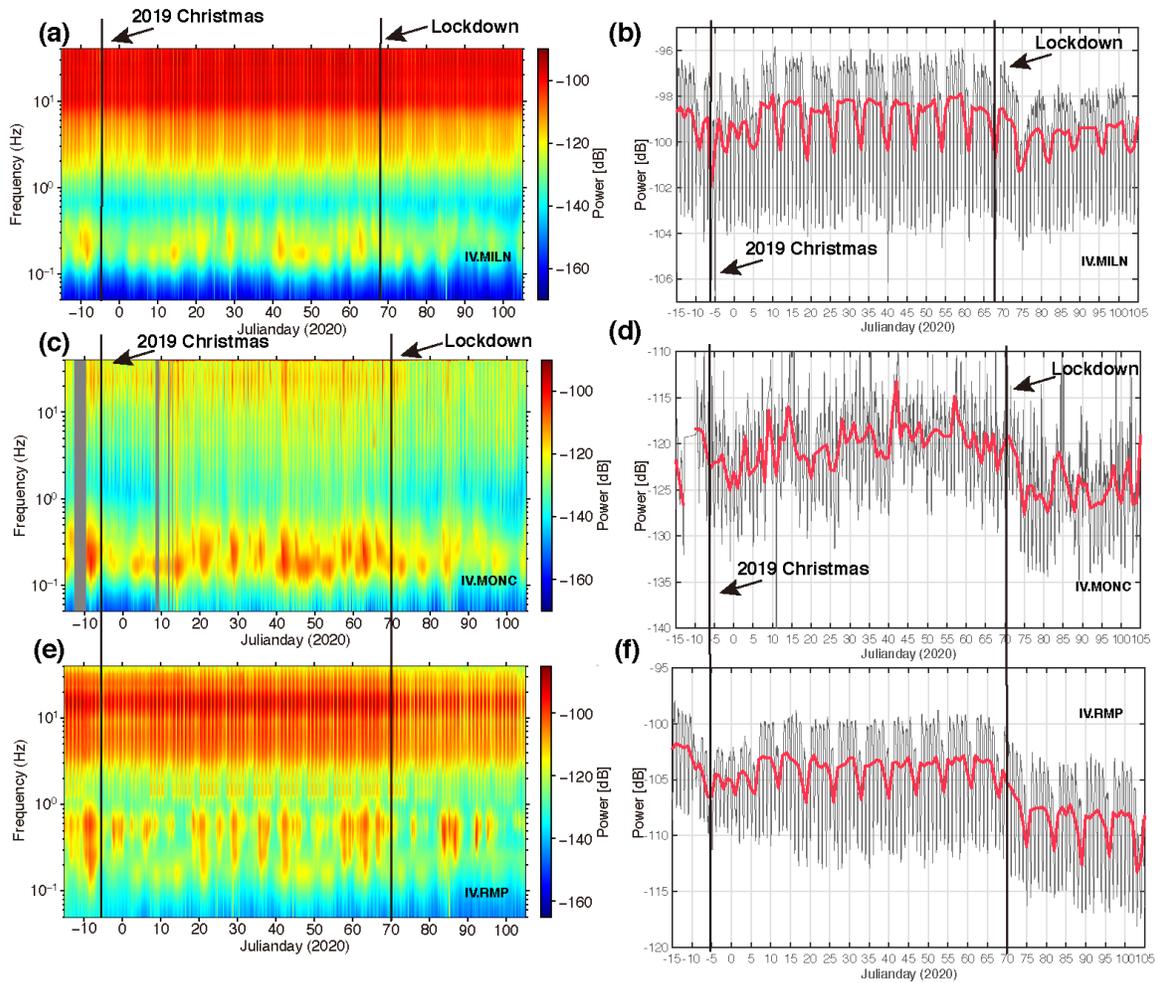

Figure 8. The energy variations of vertical-component power spectral densities at stations (a, b) IV.MILN in Milan, Italy, and (c, d) IV.MONC in the area of Torino, Italy and (e, f) IV.RMP located in ~20 km southeast of Rome, Italy. The power is measured from the vertical component in decibels relative to the ground acceleration with units of $10\log_{10}(\frac{\frac{m^2}{s^4}}{Hz})$. The left panels show noise energy as a function of time and frequency. The black lines indicate the times that the cities where the seismic stations are located went under lockdown. The right plots show the noise variations in the frequency band 10-40 Hz. The grey line is plotted in half-hour bins and the red line is plotted in one-day average bins.

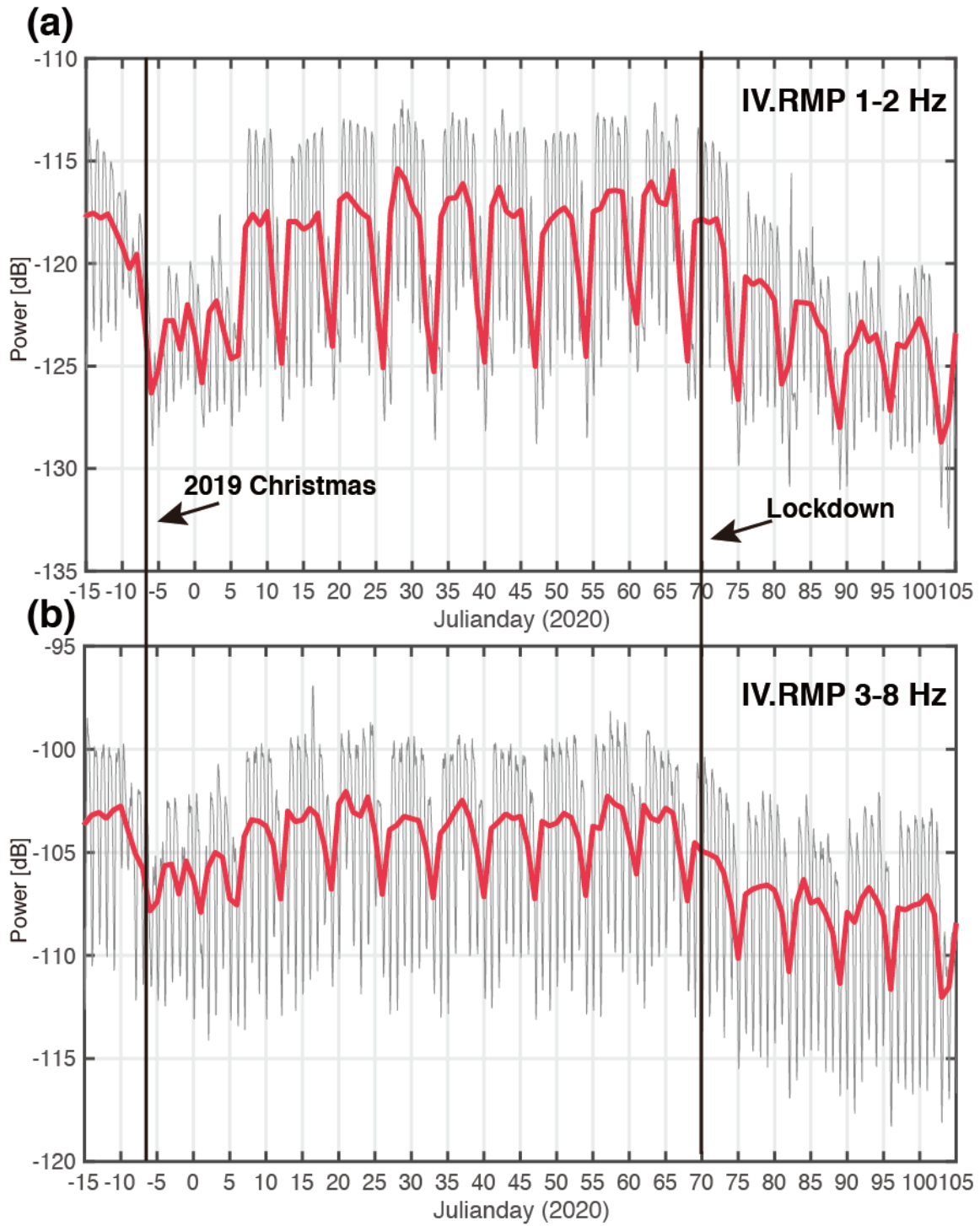

Figure 9. The energy variations of vertical-component power spectral densities at station IV.RMP (Rome, Italy) for (a) 1-2 Hz, and (b) 3-8 Hz. The power is measured from the vertical component in decibels relative to the ground acceleration with units of $10\log_{10}(\frac{\frac{m^2}{s^4}}{Hz})$. The grey line is plotted in half-hour bins and the red line is plotted in one-day average bins.